\documentclass{emulateapj}
\newcommand{\etal}{et\,al.}
\newcommand{\halpha}{H$\alpha$}
\newcommand{\kms}{km\,s$^{-1}$}
\newcommand{\lsim}{\raise0.3ex\hbox{$<$}\kern-0.75em{\lower0.65ex\hbox{$\sim$}}}

\newcommand{\HI}{H{\sc I}}
\newcommand{\HII}{H{\sc II}}
\newcommand{\msun}{M$_{\odot}$}
\newcommand{\ks}{K$_{\rm S}$}
\begin{document}
\slugcomment{The Astronomical Journal, in press}
 
\title{The Neutral Gas Dynamics of the Nearby Magellanic Irregular
  Galaxy UGCA 105}

\author{John M. Cannon\altaffilmark{1},
Elijah Z. Bernstein-Cooper\altaffilmark{1},
Ian M. Cave\altaffilmark{1},
Jon B. Harris\altaffilmark{1},
Melissa V. Marshall\altaffilmark{1},
Jacob M. Moen\altaffilmark{1},
Samilee J. Moody\altaffilmark{1},
Erin M. O'Leary\altaffilmark{1},
Stephen A. Pardy\altaffilmark{1},
Clara M. Thomann\altaffilmark{1}}

\altaffiltext{1}{Department of Physics \& Astronomy, Macalester College, 
1600 Grand Avenue, Saint Paul, MN 55105; jcannon@macalester.edu}

\begin{abstract}

We present new low-resolution \HI\ spectral line imaging, obtained
with the {\it Karl G. Jansky Very Large Array} ({\it JVLA}), of the
star-forming Magellanic irregular galaxy UGCA\,105.  This nearby (D =
3.39$\pm$0.25 Mpc), low mass (M$_{\rm HI}=$
4.3$\pm$0.5\,$\times$\,10$^{8}$ \msun) system harbors a large neutral
gas disk (\HI\ radius $\sim$ 7.2 kpc at the N$_{\rm HI}$ = 10$^{20}$
cm$^{-2}$ level) that is roughly twice as large as the stellar disk at
the B-band R$_{\rm 25}$ isophote.  We explore the neutral gas dynamics
of this system, fitting tilted ring models in order to extract a
well-sampled rotation curve.  The rotation velocity rises in the inner
disk, flattens at 72$\pm$3 \kms, and remains flat to the last measured
point of the disk ($\sim$7.5 kpc).  The dynamical mass of UGCA\,105 at
this outermost point, (9$\pm$2)\,$\times$\,10$^{9}$ \msun, is $\sim$10
times as large as the luminous baryonic components (neutral atomic gas
and stars).  The proximity and favorable inclination (55\degr) of
UGCA\,105 make it a promising target for high resolution studies of
both star formation and rotational dynamics in a nearby low-mass
galaxy.

\end{abstract}						


\section{Introduction}
\label{S1}

Dwarf galaxies offer an opportunity to study various processes that
bear on galaxy evolution.  Nearby systems allow an exploration of the
interplay between ongoing star formation and the multi-phase
interstellar medium (ISM).  Further, nearby gas-rich systems are
amenable to detailed studies of galactic rotational dynamics in the
absence of differential shear.  Many dwarfs display solid-body
rotation that is well-suited to precision rotation curve work
\citep[e.g.,][]{oh08}.  Most nearby systems appear to be dark-matter
dominated \citep{mateo98}, making them important laboratories for
studying both the luminous and the dark mass components in galaxies.

UGCA\,105 (see Table~\ref{t1} for representative qualities) is a
Magellanic-type irregular galaxy with ongoing star formation [recent
star formation as traced by the \halpha\ emission line has been
studied by \citet{kennicutt08}, \citet{lee09}, and
\citet{karachentsev10}; while the total \halpha\ luminosities differ
slightly between these works, each finds a significant \halpha-based
ongoing star formation rate of $\sim$0.06-0.07 \msun\,yr$^{-1}$].  Its
relative proximity makes it well-suited for detailed studies of the
ISM.  Using the magnitudes of the brightest stars, \citet{tikhonov92}
and \citet{karachentsev97} estimated distances of 3.2-3.3 Mpc.
Subsequent observations with the {\it Hubble Space Telescope} ({\it
  HST}) provided a distance based on the magnitude of the tip of the
red giant branch (TRGB; M$_{I}$ = $-$4.05$\pm$0.02, with little
dependence on metallicity; see {Rizzi \etal\ 2007}\nocite{rizzi07} and
references therein) of 3.15$\pm$0.32 Mpc \citep{karachentsev02}.
Subsequent analyses by \citet{jacobs09} and by the authors of the
Extragalactic Distance Database ({Tully \etal\ 2009}\nocite{tully09};
B. Jacobs, private communication) revise this slightly upward to
3.39$\pm$0.25 Mpc.  We adopt this distance measurement throughout the
present work.  At 3.39 Mpc, 1\arcsec\ corresponds to 16.4 pc.

As Figure~\ref{figcap1} shows, the stellar disk has an irregular
morphology and harbors numerous high surface brightness \HII\ regions
and widespread diffuse \halpha\ emission.  Despite these
characteristics, UGCA\,105 has by comparison remained poorly studied
in the literature.  The low Galactic latitude of the system
(13.7\degr) and the significant foreground extinction values
(E(B$-$V)=0.351 mag, or 1.51 mag of extinction in the B-band; see
discussion in footnotes to Table~\ref{t1}) may have conspired to keep
this system out of many mainstream local galaxy surveys.  As we show
in this work, the stellar and gaseous components of UGCA\,105 contain
rich morphological and kinematic structure.  This work presents the
first detailed study of the neutral gas dynamics of this nearby dwarf
galaxy.

\section{Observations and Data Handling}
\label{S2}

\HI\ spectral line imaging of UGCA\,105 was obtained with the National
Radio Astronomy Observatory's {\it Karl G. Jansky Very Large Array}
({\it JVLA}\footnote{The National Radio Astronomy Observatory is a
  facility of the National Science Foundation operated under
  cooperative agreement by Associated Universities, Inc.}) on October
10, 2011.  These data were acquired under the auspices of the
``Observing for University Classes'' program for an upper-level
astronomical techniques class at Macalester College.  The program was
officially classified as a ``Demonstration Science'' program and
carries the legacy program identification of TDEM0015.

The observations were acquired during a two-hour observation block
with the observatory in the D (most compact) configuration.  The WIDAR
correlator was used in a standard Open Shared Risk Observing single
polarization mode, providing 128 channels over a 2 MHz wide bandwidth.
The resulting channel separation of 15.625 kHz\,ch$^{-1}$ corresponds
to 3.3 km\,s$^{-1}$\,ch$^{-1}$ at the rest frequency of the
\HI\ spectral line.  Reductions followed standard prescriptions and
used both the Astronomical Image Processing System (AIPS) and the
Common Astronomy Software Applications (CASA) packages.

\begin{deluxetable*}{lcc}  
\tablecaption{Basic Characteristics of UGCA\,105} 
\tablewidth{0pt}  
\tablehead{ 
\colhead{Parameter} &\colhead{Value}}    
\startdata      
Right ascension (J2000)          &05$^{\rm h}$ 14$^{\rm m}$ 15.$^{\rm s}$3\\        
Declination (J2000)              &+62\arcdeg 34\arcmin 48\arcsec\\    
Adopted distance (Mpc)           &3.39\tablenotemark{a}\\
E(B$-$V) (Mag.)                      &0.351\tablenotemark{b}\\
M$_{\rm B}$ (Mag.)                &$-$14.70\tablenotemark{c}\\
Interferometric S$_{\rm HI}$ (Jy km\,s$^{-1}$)           &159.9\,$\pm$\,16\\
Single-dish S$_{\rm HI}$ (Jy km\,s$^{-1}$)          &163.45\tablenotemark{d}\\
\HI\ mass M$_{\rm HI}$ (\msun)   &(4.3$\pm$\,0.5)\,$\times$\,10$^8$\\
\HI\ major axis diameter (kpc)    &$\sim$14.4\tablenotemark{e}\\
B-band R$_{\rm 25}$ (kpc)        &7.5\tablenotemark{f}\\  
\enddata     
\label{t1}
\begin{small}
\tablenotetext{a}{\citet{jacobs09,tully09}} \tablenotetext{b}{From the
  Extragalactic Distance Database ({Jacobs
    \etal\ 2009}\nocite{jacobs09}, {Tully
    \etal\ 2009}\nocite{tully09}).  Note that {Schlegel
    \etal\ 1998}\nocite{schlegel98} find E(B$-$V)=0.313, but that this
  value is revised upward based on the properties of resolved stars
  (B. Jacobs, private communication).}
\tablenotetext{c}{Calculated using the observed m$_{\rm B}$=14.46\,$\pm$\,0.22
from \citet{kennicutt08}, the distance value from this table, and assuming
that the foreground Galactic reddening produces A$_{\rm B}$=1.51 magnitudes of 
extinction.}
\tablenotetext{d}{\citet{springob05}; note that those authors apply a
  $\sim$24\% correction for pointing and \HI\ self-absorption, raising the 
flux integral to 203.34\,$\pm$\,21.07 Jy km\,s$^{-1}$.}
\tablenotetext{e}{Measured as the major axis diameter at the N$_{\rm HI}$ = 10$^{20}$
cm$^{-2}$ level.}
\tablenotetext{f}{Measured as the B-band major axis diameter at the 25.0 mag\,arcsec$^{-2}$ isophote from \citet{devaucouleurs91}.}
\end{small}
\end{deluxetable*}   

UGCA\,105 has a low systemic velocity (V$_{\rm sys}$ = 90.8$\pm$2.0
km\,s$^{-1}$, derived from the tilted ring analysis discussed in
\S~\ref{S3.2}) for its distance well outside the Local Group, and
hence \HI\ emission from UGCA\,105 is relatively close in velocity
space to the foreground emission from the Milky Way.  The selection of
the band center frequency and the 2 MHz bandwidth was adequate to
cleanly separate the Milky Way and UGCA\,105 emission components.
Roughly 40 line-free channels were available at velocities higher than
those occupied by the target galaxy, and these were used to fit the
underlying continuum with a first-order polynomial and remove it from
all channels of the cube.  UGCA\,105 occupies roughly 54 channels of
the cube; Milky Way contamination is severe in the $\sim$25
lowest-velocity channels, and about 8 channels separate UGCA\,105 and
the foreground components.  To be cautious, we do not use these 8
line-free channels in the continuum removal.  We conclude that
foreground Milky Way emission does not affect the portions of our
cubes that contain \HI\ spectral line emission from UGCA\,105.

\begin{figure*}
\epsscale{1.0}
\plotone{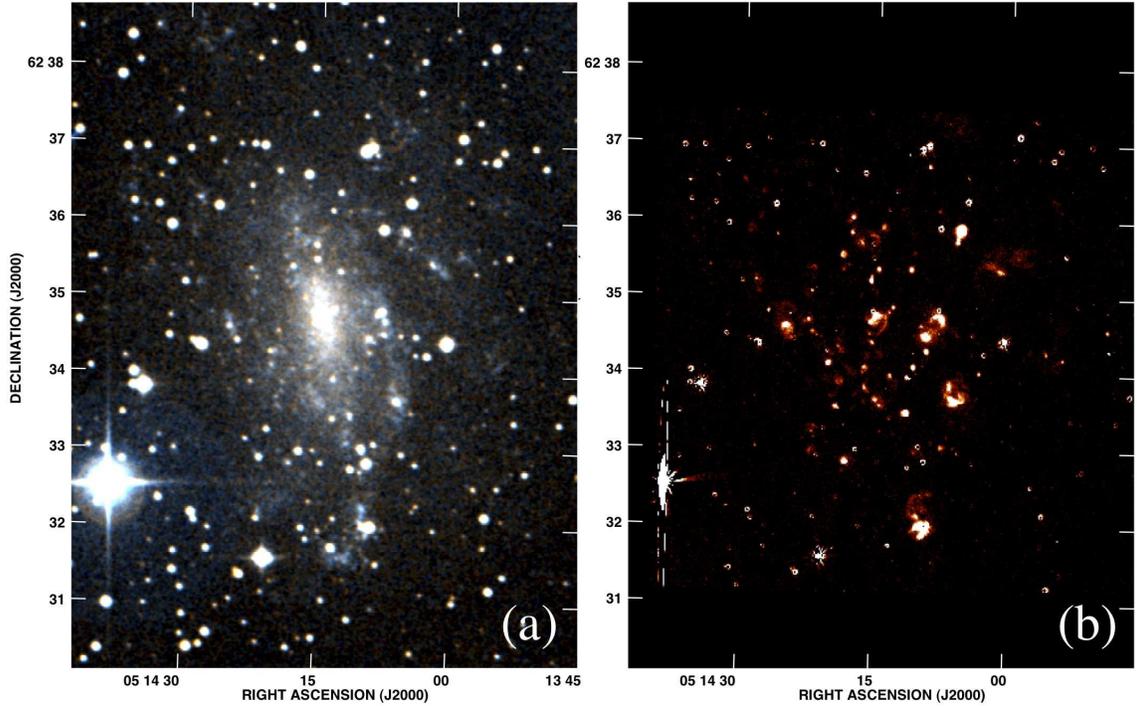}
\caption{(a) 3-color image of UGCA\,105 created from B, V, and R band
  {\it Digitized Sky Survey} images.  The faint arc of blue emission
  in the upper right corner is scattered light from a bright
  foreground star off the field of view.  (b) Continuum-subtracted
  H$\alpha$ image of UGCA\,105 (image courtesy Janice Lee and Robert
  C. Kennicutt, Jr.).  There are multiple star formation regions
  throughout the disk of UGCA\,105.  The field of view shown in these
  images is smaller than that shown in Figure~\ref{figcap6}.}
\label{figcap1}
\end{figure*}

The final spectral line cube used in this analysis was produced using
the AIPS IMAGR task with a ROBUST weighting of 0.5; this cube was
cleaned to the rms level (0.003 Jy\,beam$^{-1}$) found in 2-3 line-free
channels (purposefully included in the cube and not including emission
from UGCA\,105).  This original cube has a resolution element of
53.58\arcsec$\times$\,42.01\arcsec.  To produce the moment maps used
below, we follow procedures similar to those described in
\citet{cannon09,cannon10,cannon11}.  The full-resolution spectral line
cube is convolved to a circular beam size of 60\arcsec.  The rms noise
level in the line-free channels of this convolved cube is found to be
0.0025 Jy\,beam$^{-1}$.  The convolved cube is blanked at 2.5 times this
rms level (0.00625 Jy\,beam$^{-1}$), and the resulting blanked cube is
then carefully examined by hand to identify regions of ``real''
emission that are present in three or more consecutive channels.  This
blanked cube is then used as a template to blank the original cube
when convolved to the smallest possible circular beam size.  Thus, the
final data cube produced here has a 54\arcsec\ circular beam (physical
resolution element $=$890 pc).  The noise in the final cube is 0.0047
Jy\,beam$^{-1}$.

The first two moment maps at 54\arcsec\ resolution (representing
integrated intensity and intensity weighted velocity) were created
from this data cube using the XMOM task in AIPS.  The final moment
zero image was converted to column density units (atoms cm$^{-2}$)
using standard relations.  Edge effects in the final moment maps were
minimized by blanking the final moment zero image below the
1\,$\times$\,10$^{20}$ cm$^{-2}$ level, and using this same level for
all moment maps.  The images of integrated \HI\ emission are discussed
in detail in \S~\ref{S3}.

The total \HI\ flux integral, derived from the 54\arcsec\ cube, is
159.9\,$\pm$\,16 Jy\,\kms; this is in excellent agreement with the
single-dish flux integral of 163.45 Jy\,\kms\ found by
{Springob \etal\ (2005}\nocite{springob05}; however, those authors do
apply an upward $\sim$24\% correction for \HI\ self-absorption,
bringing their absorption-corrected flux integral to
203.34\,$\pm$\,21.07 Jy\,\kms).  Using the flux integral derived from
our {\it JVLA} imaging, we derive a total \HI\ mass of
(4.3$\pm$0.5)\,$\times$\,10$^{8}$ \msun.

During the reduction of the \HI\ spectral line data, we averaged 28
line-free channels to produce a 1.4 GHz radio continuum image.  This
image has a beam size of 53.6\arcsec\,$\times$\,41.6\arcsec\ and an
rms noise level of 6.5\,$\times$\,10$^{-4}$ Jy\,beam$^{-1}$. There is
weak positive point source emission detected in the southern part of
the disk of UGCA\,105.  This source is catalogued in NVSS images
\citep{condon98} as NVSS\,J051418+623136 with a flux density of
2.1\,$\times$\,10$^{-2}$ Jy; we measure a flux density of
(1.6\,$\pm$0.5)\,$\times$\,10$^{-2}$ Jy in our continuum image.  This
object is most likely a background continuum source, such as a QSO,
and is not coincident with any high surface brightness
\halpha\ emission from UGCA\,105 (see Figure~\ref{figcap1}).  The
5$\sigma$ upper limit to the global 1.4 GHz radio continuum flux
density from UGCA\,105 is derived to be S$_{\rm 1.4 GHz}
\lsim$3.2\,$\times$\,10$^{-3}$ Jy.  This flux density is comparable to
the total thermal radio continuum flux expected based on the strength
of the integrated \halpha\ emission ({Kennicutt
  \etal\ 2008}\nocite{kennicutt08} publish a total
\halpha\,$+$\,[N~II] flux of 2.5\,$\times$\,10$^{-12}$
erg\,s$^{-1}$\,cm$^{-2}$) using the prescriptions in {Caplan \&
  Deharveng (1986)}\nocite{caplan86}.  Given these limitations, and
the distributed star formation throughout the disk of UGCA\,105 (see
Figure~\ref{figcap1}), we do not study the detailed nature of the
radio continuum emission in UGCA\,105 with the present data.

\section{Gaseous, Stellar, and Dark Components}
\label{S3}
\subsection{The Neutral Gas Distribution and Dynamics}
\label{S3.1}

\begin{figure*}
\epsscale{1.0}
\plotone{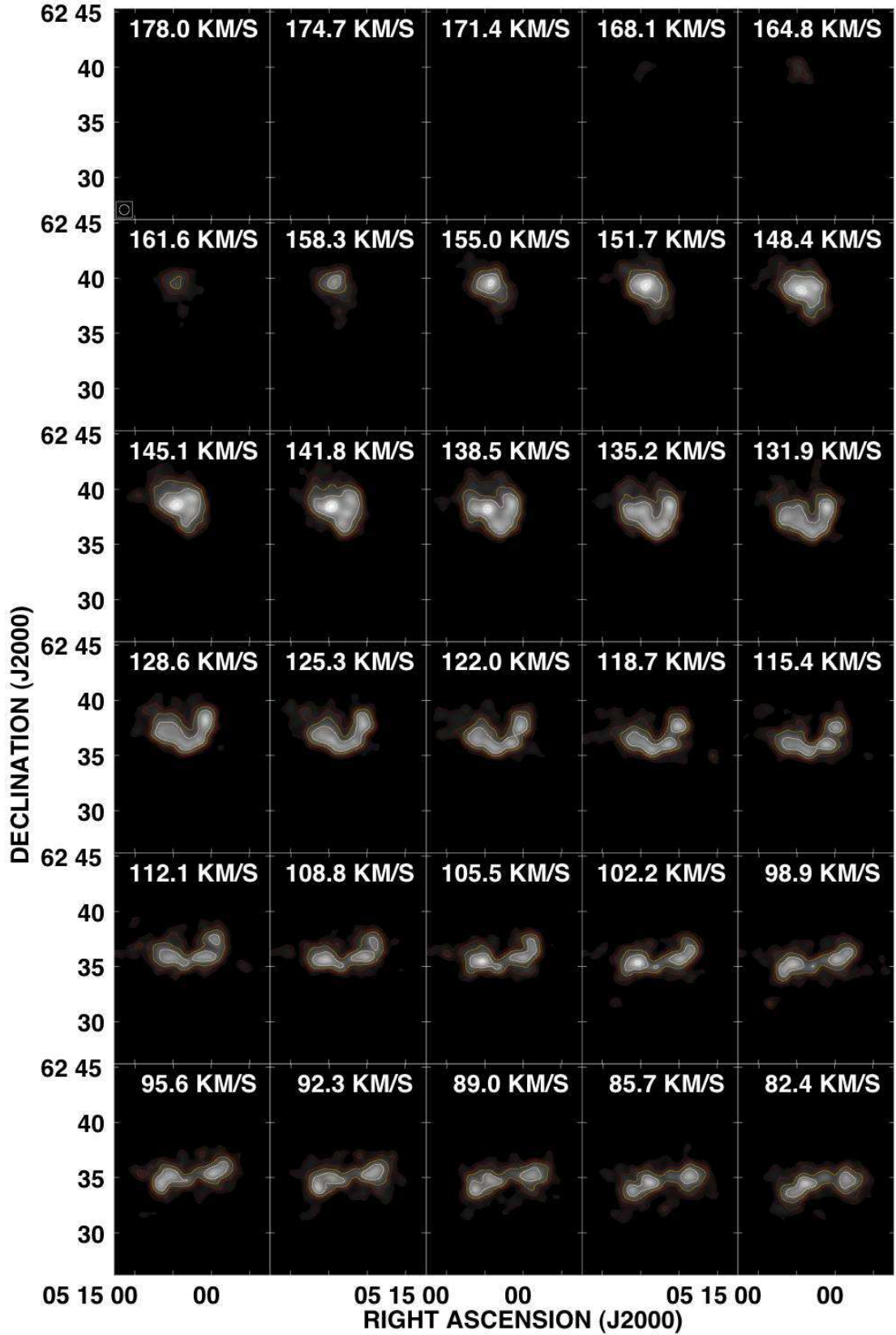}
\caption{30 channel maps from the 54\arcsec\ resolution datacube;
  heliocentric velocity is noted in each frame, and the beam size is
  shown in the top left panel.  Contours are at levels of
  (3,6,12,24,48)\,$\sigma$, where $\sigma$=2.5\,$\times$\,10$^{-3}$
  Jy\,beam$^{-1}$ is the rms noise measured in line-free channels of
  the cube.}
\label{figcap2}
\end{figure*}

\begin{figure*}
\epsscale{1.0}
\plotone{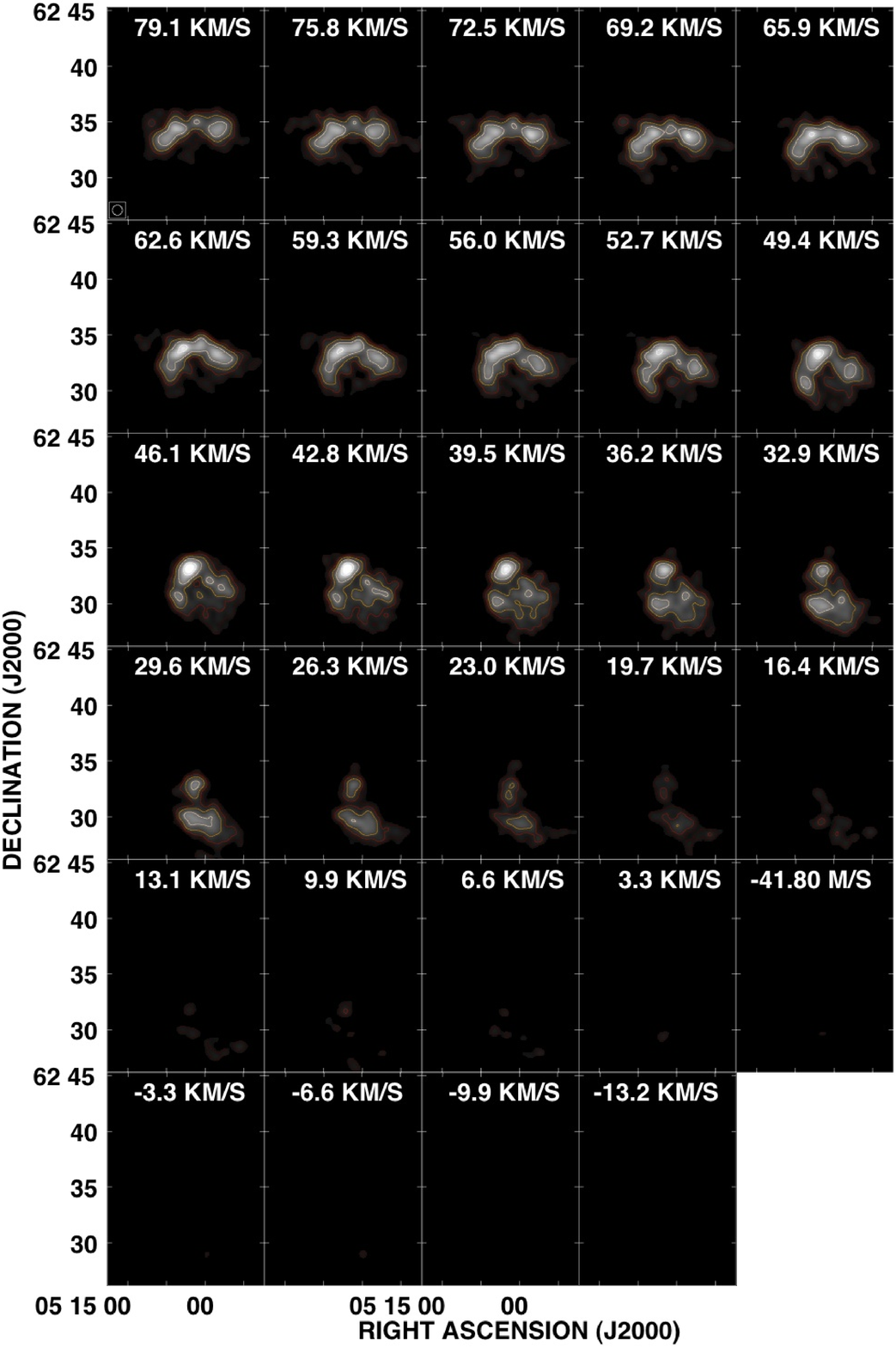}
\caption{Same as Figure~\ref{figcap2}, but for the remaining 29
  channels of the 54\arcsec\ resolution datacube.}
\label{figcap3}
\end{figure*}

\HI\ emission from UGCA\,105 is detected at high significance in 54
channels spanning the heliocentric radial velocity range of 168.1 --
$-$6.6 \kms.  Figures~\ref{figcap2} and \ref{figcap3} show the
individual channel maps of the 54\arcsec-resolution datacube.  The
ordered rotation of the system is prominent as a classical ``butterfly
diagram'' moving from southwest to northeast with increasing
heliocentric velocity.  The \HI\ surface brightness is slightly lower
in the southwest region of the disk compared to the northeast,
although the bulk kinematics are very well defined throughout the
disk.

\begin{figure*}
\epsscale{1.0}
\plotone{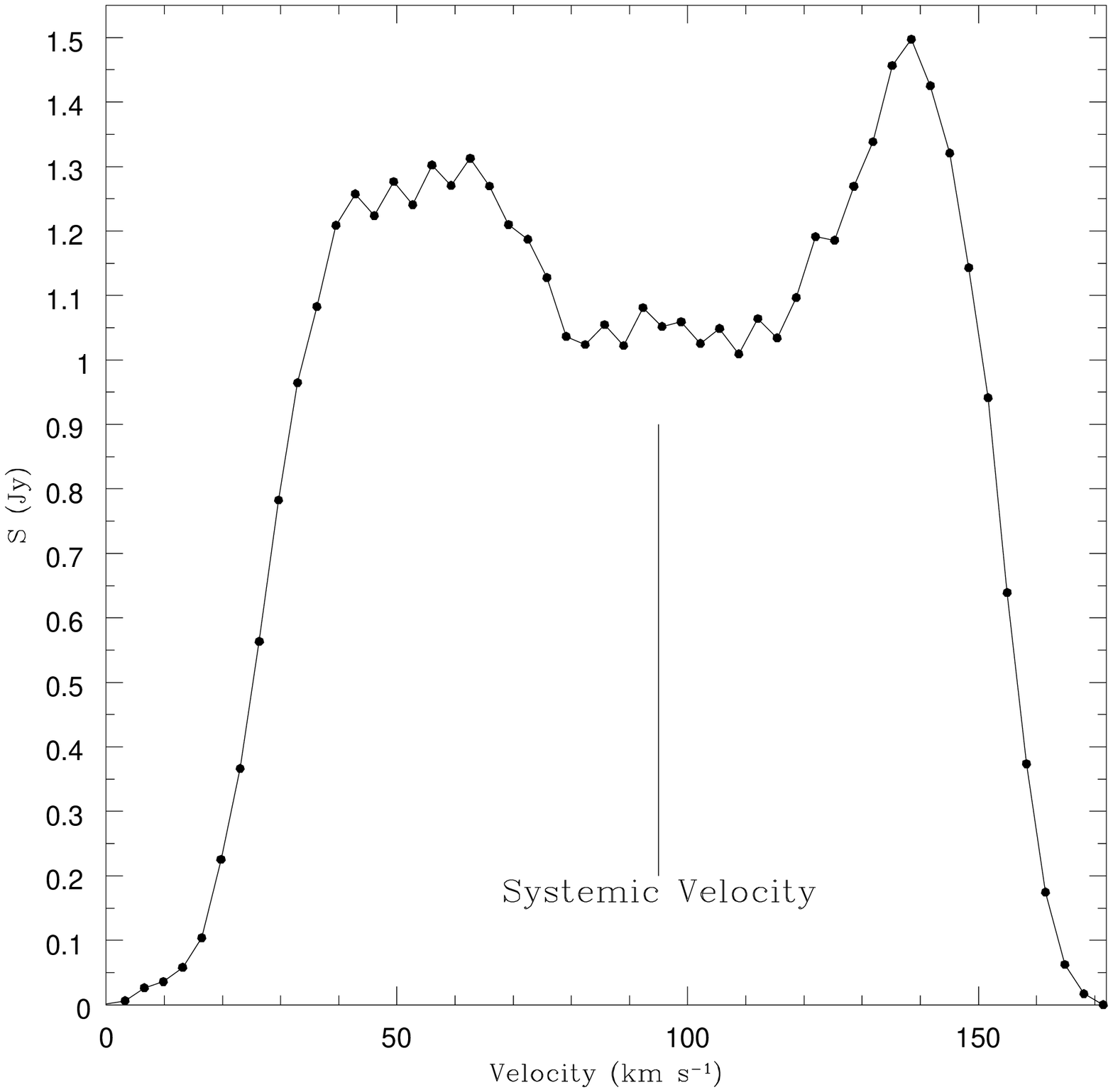}
\caption{Global \HI\ profile of UGCA\,105, created by summing the flux
  in each channel of the 54\arcsec\ resolution blanked datacube. The
  systemic velocity, derived from our tilted ring analyses (see
  Section 3.1), is 90.8$\pm$2.0 \kms.}
\label{figcap4}
\end{figure*}

We create the global \HI\ profile shown in Figure~\ref{figcap4} by
summing the flux in each of the channel maps shown in
Figures~\ref{figcap2} and \ref{figcap3}.  As expected from the
``butterfly diagram'' noted above, we see a classic double-horn
\HI\ line profile.  The higher velocity (northeast) region of the disk
shows a slightly higher integrated flux per channel.  We fit this
profile to derive the systemic velocity of the system (95\,$\pm$\,5
\kms).  We refine this estimate in our kinematic analysis (see below)
to V$_{\rm sys}$ = 90.8\,$\pm$\,2 \kms, and show V$_{\rm sys}$ as a
vertical line in Figure~\ref{figcap4}.

\begin{figure*}
\plotone{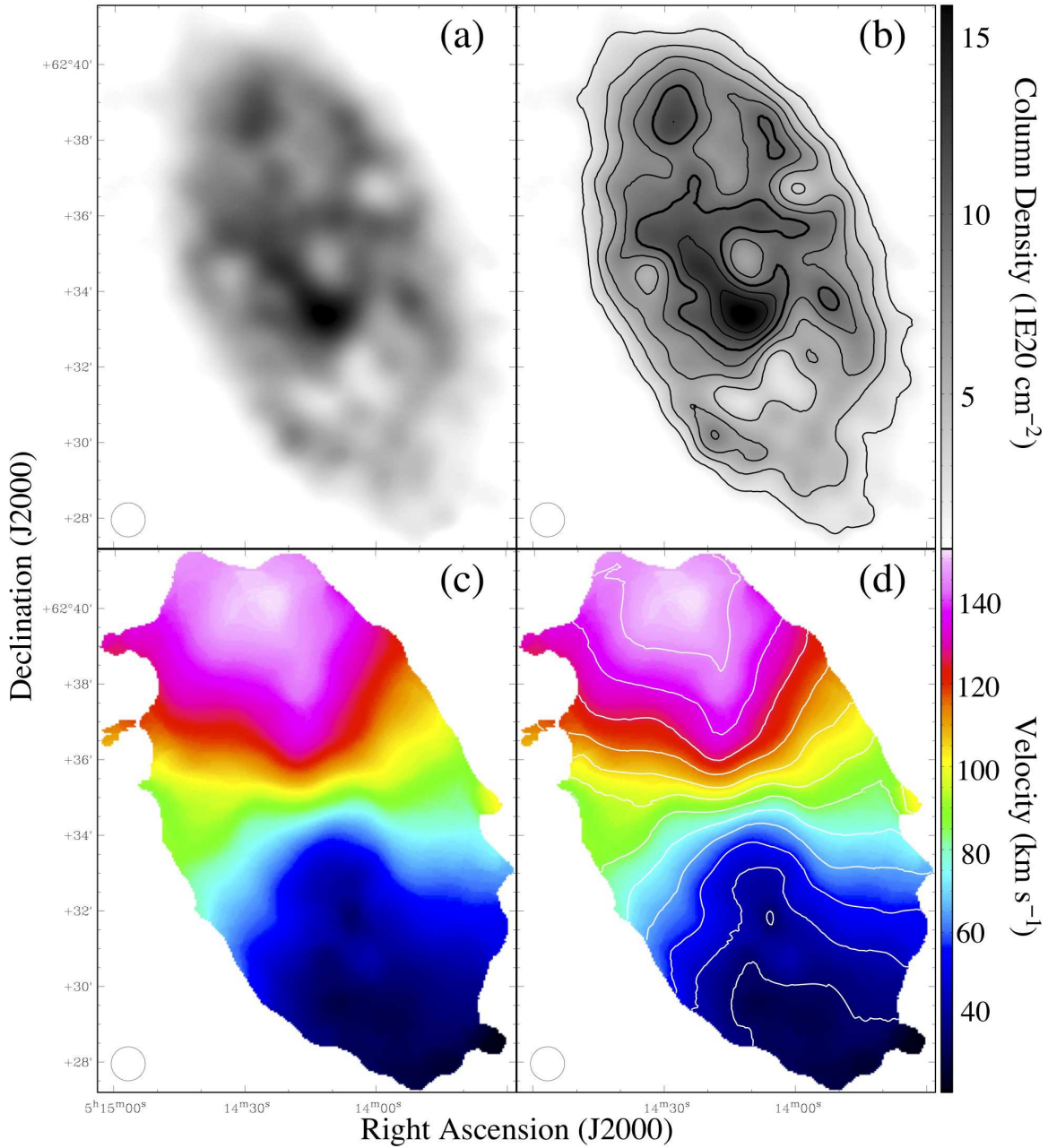}
\caption{Comparison of the \HI\ column density distribution ({\it a,
    b}) and the integrated velocity field ({\it c, d}); the beam size
  is 54\arcsec.  The contours in {\it b} are at the (2, 4, 6, 8, 10,
  12, 14, 16)\,$\times$\,10$^{21}$ cm$^{-2}$ level.  The contours in
  ({\it d}) show velocities between 35 and 145 \kms, separated by 10
  \kms.  UGCA\,105 is undergoing well-ordered rotation throughout most
  of the disk.}
\label{figcap5}
\end{figure*}

Figure~\ref{figcap5} shows two-dimensional images of the
\HI\ morphology and kinematics.  The moment 0 (integrated
\HI\ intensity) images show rich morphological structure in the
neutral gas disk.  There is low surface brightness \HI\ gas in the
outer disk, but throughout most of the inner disk the column densities
rise above 10$^{21}$ cm$^{-2}$ at 54\arcsec\ (890 pc) resolution.
There are regions of the inner disk with comparatively low \HI\ column
densities that are roughly the size of the synthesized beam.  While
these features are suggestive of \HI\ holes and shells, we await
higher spatial resolution imaging in order to study these features in
detail.


As expected from the well-behaved channel maps shown in
Figures~\ref{figcap2} and \ref{figcap3}, the intensity weighted
velocity field shown in Figure~\ref{figcap5} is very symmetric at this
physical resolution.  Within the inner disk the isovelocity contours
are mostly parallel and the major axis of rotation is well-defined
(position angle $\sim$15\degr\ east of north; see more detailed
discussion below). The bulk gas kinematics are well-behaved and
amenable to detailed rotation curve analysis (see next subsection).

\begin{figure*}
\plotone{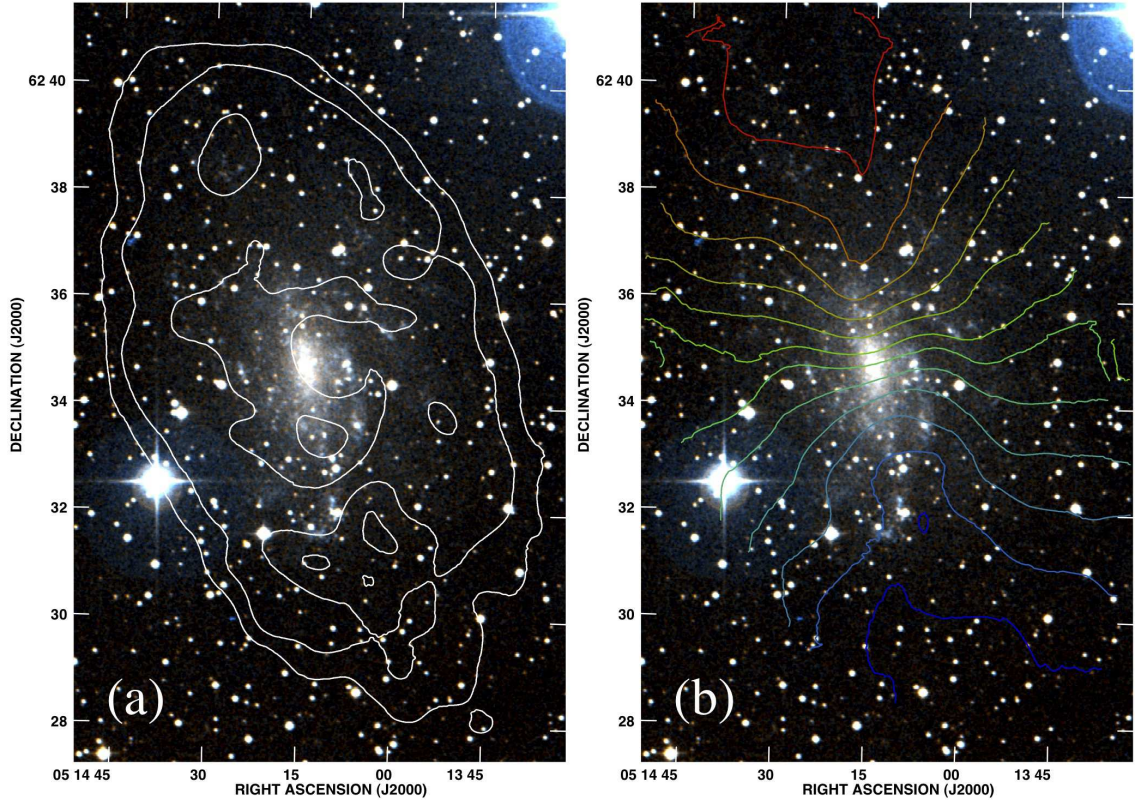}
\caption{{\it Digitized Sky Survey} 3-color optical images, overlaid
  with contours showing HI column density (a) and intensity weighted
  velocity (b).  The contours in (a) and (b) are at levels of (2.5, 5,
  10, 15)\,$\times$\,10$^{20}$ cm$^{-2}$ and (35, 45, 55, 65, 75, 87,
  95, 105, 115, 125, 135, 145) km\,s$^{-1}$, respectively.  Most of
  the high surface brightness stellar disk is located within the dense
  inner gaseous disk (N$_{\rm HI}$ $>$ 10$^{21}$ cm$^{-2}$).}
\label{figcap6}
\end{figure*}

The neutral gas disk (\HI\ radius $\sim$ 7.2 kpc at the N$_{\rm HI}$ =
10$^{20}$ cm$^{-2}$ level) is roughly twice as large as the stellar
disk at the B-band R$_{\rm 25}$ isophote (see Table~\ref{t1}).  To
allow a straightforward comparison of the stellar and gaseous disk
components, we present in Figure~\ref{figcap6}(a) an optical
three-color image (created from {\it Digitized Sky Survey} images)
overlaid with contours of \HI\ column density. The high surface
brightness stellar disk is cospatial with the dense inner regions of
the gaseous disk (N$_{\rm HI}$ $>$ 10$^{21}$ cm$^{-2}$).
Figure~\ref{figcap1} verifies that most of the ongoing star formation
in UGCA\,105 is also concentrated within this inner disk (see further
discussion below).  

Figure~\ref{figcap6}(b) shows the same three color image overlaid with
color-coded isovelocity contours.  As expected, the high surface
brightness stellar disk is located within the innermost and
well-behaved section of the intensity weighted velocity field.  The
extent of the well-sampled velocity field compared to the stellar
disk, and its coherent rotation at large galactocentric radii,
immediately imply that there is a substantial mass component in the
outer disk regions.  We explore this in more detail in the next
subsections.

\begin{figure*}
\plotone{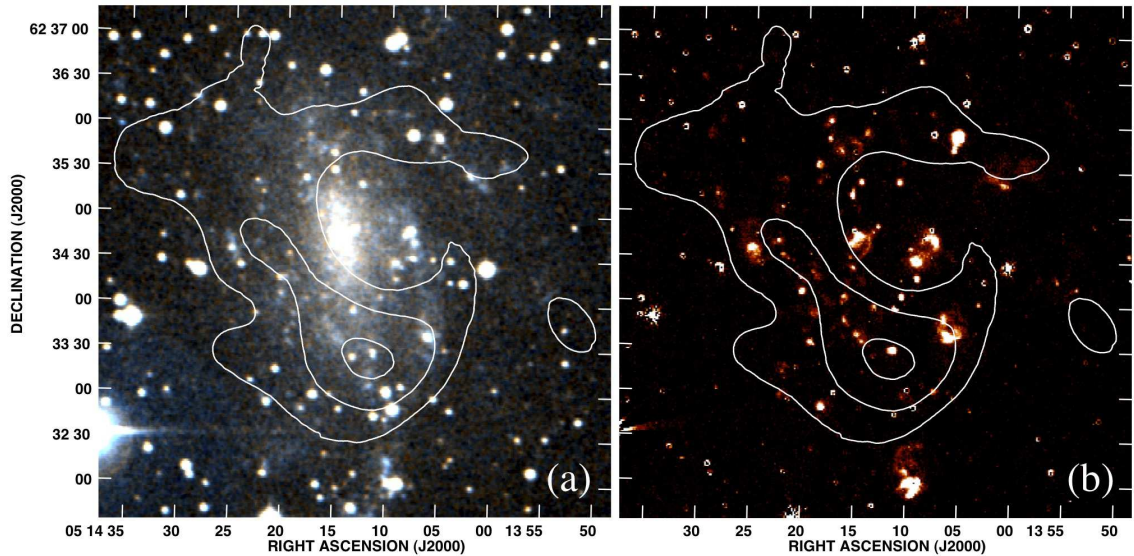}
\caption{{\it Digitized Sky Survey} 3-color optical image (a) and
  continuum-subtracted H$\alpha$ image (b) of the inner disk of
  UGCA\,105, overlaid with \HI\ column density contours at the 
(10, 12.5, 15)\,$\times$\,10$^{20}$ cm$^{-2}$ levels.  Most, but not all, 
of the ongoing star formation in UGCA\,105 is located in the regions of 
highest neutral gas surface density.  A detailed comparative study of the 
neutral gas properties and the ongoing star formation requires higher spatial 
resolution data.}
\label{figcap7}
\end{figure*}

At the spatial resolution of these data ($\sim$890 pc) we can perform
only bulk comparisons of the local \HI\ surface density with regions
of ongoing star formation.  Such a first order comparison is shown in
Figure~\ref{figcap7}, where \HI\ column density contours at the
(10, 12.5, 15)\,$\times$\,10$^{20}$ cm$^{-2}$ levels are overlaid on
zoomed versions of the same 3-color optical and continuum-subtracted
\halpha\ images as shown in Figure~\ref{figcap1}.  Most, but not all,
of the ongoing star formation in UGCA\,105 is coincident with the
highest surface density regions of the neutral gas disk.  A detailed
comparative study of the local neutral gas surface density and the ongoing 
star formation in UGCA\,105 would be very fruitful, and we defer such discussion
until higher spatial resolution \HI\ imaging is available.

\subsection{Dynamical Analysis of the UGCA\,105 Neutral Gas Disk}
\label{S3.2}

\begin{figure*}
\plotone{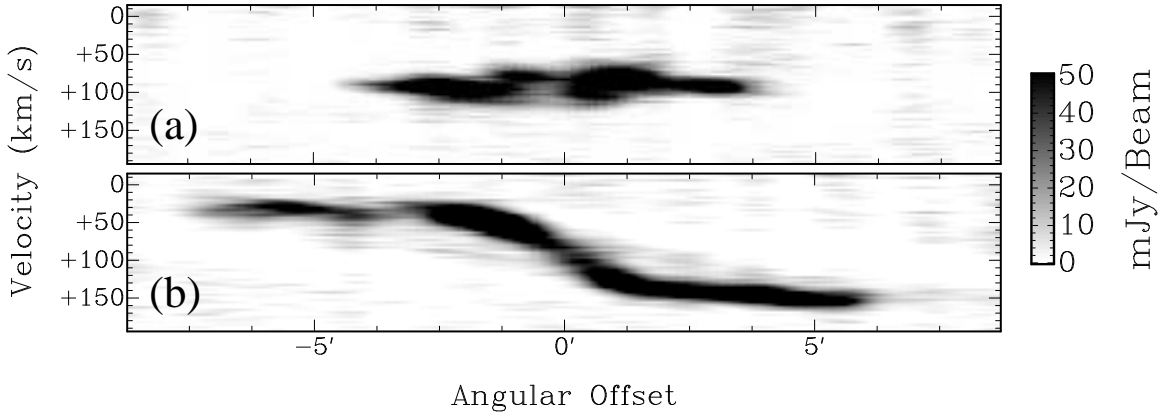}
\caption{Minor (position angle 105.12\degr, measured east of north;
  {\it a}) and major (position angle 15.12\degr, measured east of
  north; {\it b}) axis position-velocity diagrams.  Each cut is
  centered on the dynamical center of the galaxy
  (5$^{h}$14$^{m}$13.76$^{s}$, 62{\degr}34{\arcmin}53.0{\arcsec}) as
  derived from the rotation curve analysis (see Figure~\ref{figcap9}
  and discussion in \S~\ref{S3.2}).}
\label{figcap8}
\end{figure*}

The proximity of UGCA\,105, and the well-ordered rotation throughout
most of the \HI\ disk, are conducive to detailed rotation curve
analysis.  A first order understanding of the bulk kinematics is
provided by a simple position velocity (PV) cut through the full
\HI\ cube.  By varying the position angle of the PV cut, one can
easily identify the direction of maximum velocity gradient in the
galaxy; this offers a simple prior on the kinematic position angle.
Similarly, the position angle of zero velocity gradient will be very
close to the kinematic minor axis.  As shown in Figure~\ref{figcap8},
the PV cut along the major axis (which is not corrected for
inclination) already shows a flat rotation profile outside of the
inner solid body disk.  Similarly, the velocity of the kinematic minor
axis is in good agreement with the systemic velocity derived above and
shows a flat profile in velocity as a function of position.


Given these first constraints on the gas kinematics, we next fitted
tilted ring models to the velocity field using tools in the {\it
  GIPSY} software package\footnote{The Groningen Image Processing
  System (GIPSY) is distributed by the Kapteyn Atronomical Institute,
  Groningen, Netherlands.}.  In an iterative sequence, we fit the
systemic velocity (V$_{\rm sys}$), dynamical center position, position
angle (P.A.), inclination ({\it i}) and rotational velocity for the
galaxy as a whole, and for the receding and approaching sides
individually.  The priors on systemic velocity (see
Figures~\ref{figcap4} and \ref{figcap8}) and on kinematic position
angle (see Figure~\ref{figcap8}) allow one to narrow the starting
points in parameter space (although checks against the same procedures
executed without the priors showed no difference in the final best-fit
result).  We experimented with all available sequences of parameter
fitting (i.e., fitting each different variable in turn throughout the
procedure); the final result is robust against any choice of fitting
sequence.

\begin{figure*}\vspace{-1 cm}
\epsscale{0.715}
\plotone{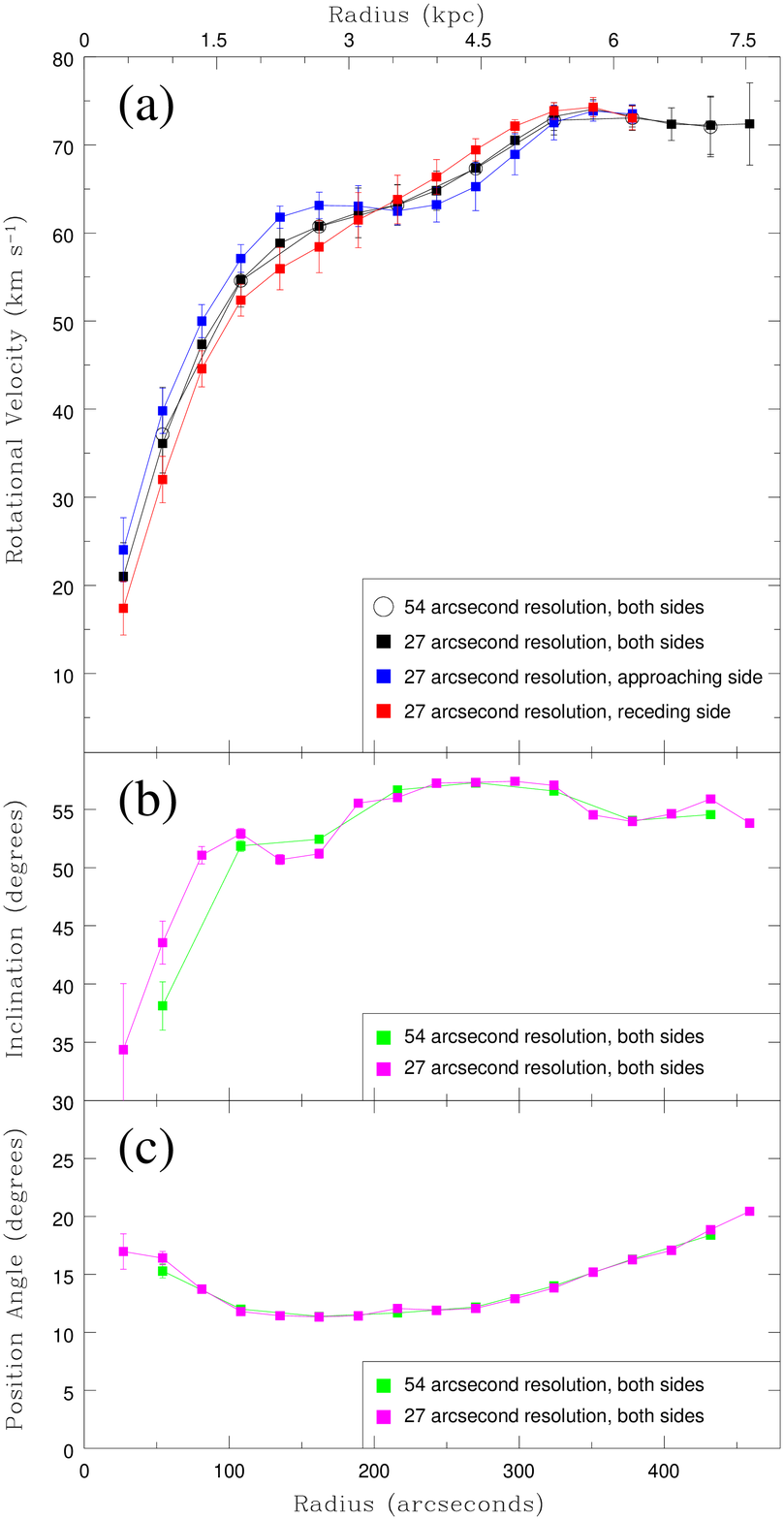}
\caption{(a) Rotation curves of UGCA\,105 derived by fitting tilted
  ring models to the intensity-weighted velocity field.  The open
  circles use annuli separated by the full beam width (54\arcsec) and
  are independent at each measured position.  The solid black points
  use annuli with 27\arcsec\ widths (i.e., half of the beam size) and
  are only technically independent at every other data point.  The
  black points are fit to both sides of the disk simultaneously.  The
  blue and red curves fit tilted rings to either the approaching
  (blue) or receding (red) sides of the disk, using the same
  parameters of V$_{\rm sys}$ (90.8 km\,s$^{-1}$), inclination
  (54.9\degr), dynamical center position (5$^{h}$14$^{m}$13.76$^{s}$,
  62{\degr}34{\arcmin}53.0{\arcsec}), and position angle (15.12\degr)
  as used for the 27\arcsec\ resolution fit to both sides of the disk.
  The lower panels show the inclination angle (b) and the kinematic
  major axis position angle (c) as functions of radius.  The profiles
  shown in (b) and (c) were derived using fixed values for V$_{\rm
    sys}$ and and dynamical center position; there are only minor
  variations of these parameters with radius.}
\label{figcap9}
\end{figure*}

We obtained optimal rotation curve fits to the observed low resolution
velocity field by fixing parameters as follows: {\it i} =
55\arcdeg$\pm$2\arcdeg; P.A. = 15\arcdeg$\pm$2\arcdeg\ (which is in
good agreement with the apparent major axis position angle of the
stellar disk; see, e.g., Figure~\ref{figcap1}); V$_{\rm sys}$ =
(90.8$\pm$2.0) \kms; and dynamical center position ($\alpha$,$\delta$,
J2000) = (5$^{h}$14$^{m}$13.76$^{s}$,
62{\degr}34{\arcmin}53.0{\arcsec}).  The resulting curves are shown at
both 27\arcsec\ and 54\arcsec\ resolution in Figure~\ref{figcap9};
note that the solid black points use annuli with 27\arcsec\ widths
(i.e., half of the beam size) and are only technically independent at
every other data point.  In Figure~\ref{figcap9} we also show the
profiles of kinematic position angle and inclination (derived using
fixed values for V$_{\rm sys}$ and dynamical center position);
there are only minor variations of these parameters with radius.  The
rotation curve clearly demonstrates solid-body rotation to
$\sim$100\arcsec\ (1.5 kpc), a more shallow rise between 1.5 and 5.0
kpc, and a flattening at $\sim$72 \kms\ to the detection limit in the
outer disk ($\sim$7.5 kpc).  Assuming circular orbits, the implied
dynamical mass at this outermost radius is
(9$\pm$2)\,$\times$\,10$^{9}$ \msun.

This dynamical mass measurement can be directly compared to the total
baryonic mass of UGCA\,105.  As noted above, our {\it JVLA} images
reveal a total \HI\ mass of (4.3$\pm$0.5)\,$\times$\,10$^{8}$ \msun.
Given the low mass of UGCA\,105, it is likely to be a metal-poor
system (see, e.g., the mass-metallicity relationship presented by
{Tremonti \etal\ 2004}\nocite{tremonti04}) whose molecular content is
difficult to measure observationally.  While \citet{leroy05} do find a
marginal detection of CO in UGCA\,105 (I$_{\rm CO}$ =
(0.61\,$\pm$\,0.16) K\,km\,s$^{-1}$), a direct measurement of the
molecular mass in this galaxy is not currently available.  Thus,
following \citet{cannon10}, we include a 35\% correction for Helium
and molecular material and adopt (5.9$\pm$0.7)\,$\times$\,10$^{8}$
\msun\ as the total gas mass (M$_{\rm gas}$) of UGCA\,105.

\begin{figure*}
\plotone{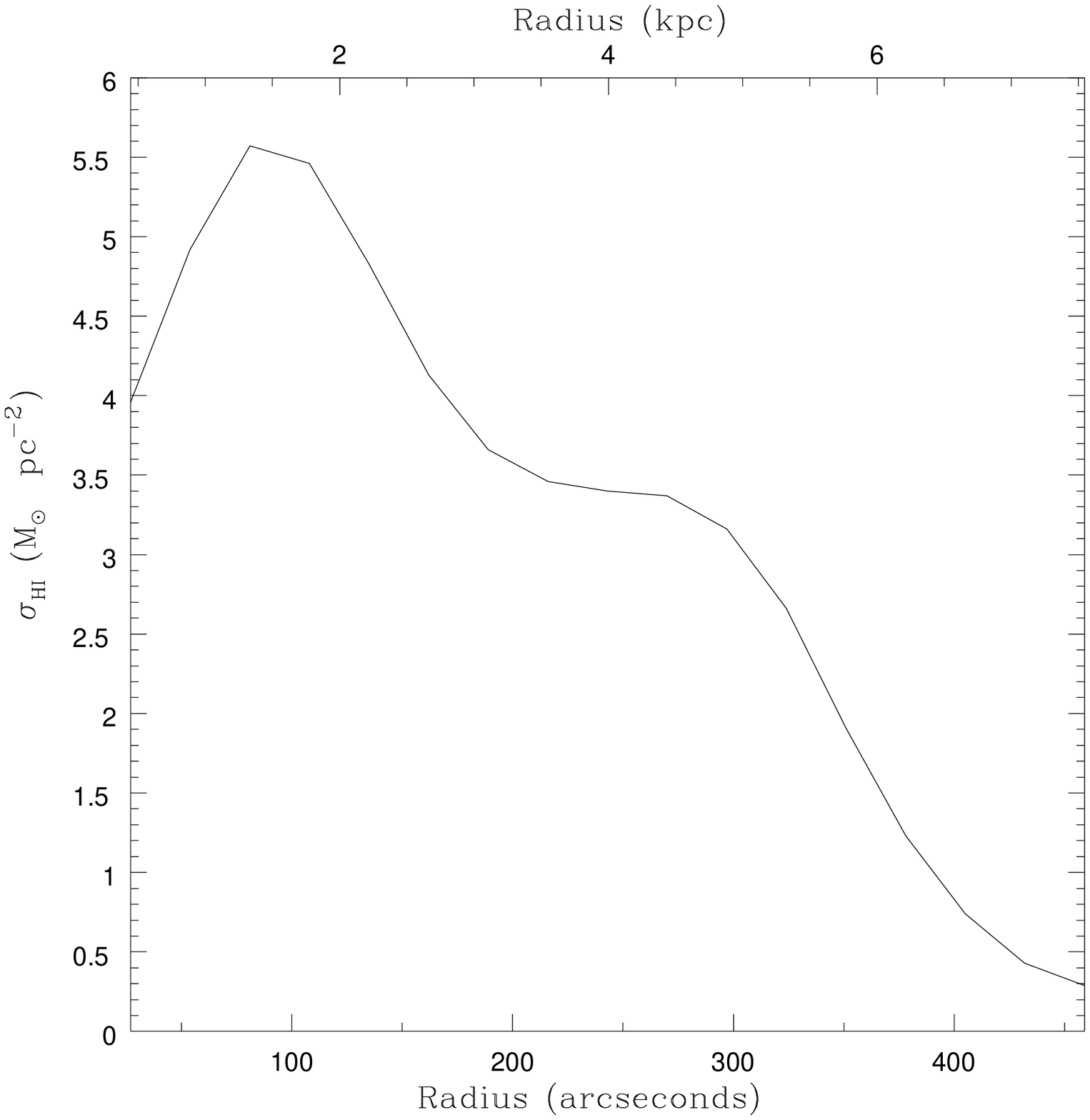}
\caption{Radially averaged \HI\ mass surface density profile of
  UGCA\,105, created by summing \HI\ emission in concentric rings
  emanating from the dynamical center found in our rotation curve
  analysis.  The solid line was created using annuli spaced by
  27\arcsec\ (half of the 54\arcsec\ circular beam size; see
  discussion in \S~\ref{S3.2} and Figure~\ref{figcap9}).}
\label{figcap10}
\end{figure*}

Our tilted ring analysis also allows us to examine the radial behavior
of \HI\ mass density per unit area throughout the gas disk.  We again
used the {\it GIPSY} software package to integrate the \HI\ flux per
unit area, in concentric rings separated by the beam widths, after
correcting for inclination and the galaxy's major axis position angle
(see Figure~\ref{figcap9} and discussion above).  The resulting plot,
shown in Figure~\ref{figcap10}, demonstrates that the innermost few
hundred pc of the \HI\ disk contains significantly less neutral gas
per unit area than the regions further out in the disk; this could be
tantalizing though marginal evidence for a molecular region in the
inner disk (indeed, the marginal CO detection by {Leroy
  \etal\ 2005}\nocite{leroy05} is in close proximity to the dynamical
center position).  The highest mass surface densities are located
between $\sim$1--2 kpc from the dynamical center.  

\subsection{Luminous and Dark Mass Components in UGCA\,105}
\label{S3.3}

The infrared luminosity of UGCA\,105 is estimated using the 2MASS
broadband infrared photometry presented in the 2MASS Extended Source
Catalog \citep{jarrett00}.  The total J, H, and \ks\ magnitudes are
12.274\,$\pm$\,0.030, 11.510\,$\pm$0.037, and 11.364\,$\pm$\,0.052,
respectively.  We assume a negligible difference between K and
\ks\ magnitudes and adopt L$_{\rm K,\,\odot} = +$3.33
\citep{cox00,bessel79}.  The total K-band luminosity of UGCA\,105 is
found to be L$_{\rm K} = $ $\sim$7.8\,$\times$\,10$^7$ L$_{\odot}$
after correcting for Milky Way extinction.

To derive the mass of the stellar component, we apply the techniques
discussed in detail in \citet{cannon10}.  Briefly, we use the models
presented in \citet{bell01}, which take as input the observed color
and luminosity of a stellar population.  The models allow one to
estimate a stellar mass-to-light ratio (M/L) assuming a range of
metallicities, initial mass functions, and galaxy evolution
properties.  Given the substantial Milky Way foreground screen, the
near-infrared colors are optimal; further, these colors are less
affected by recent or ongoing star formation within UGCA\,105 than are
optical colors.  Applying the model M/L ratios and the observed K-band
luminosity, we can then estimate the stellar mass of UGCA\,105.  Using
this approach we find that the underlying stellar mass in UGCA\,105 is
M$_{\star}$ $=$ (1.8\,$\pm$\,0.8)\,$\times$\,10$^{8}$ \msun.

The sum of the luminous baryons (gas and stars) in UGCA\,105 is
estimated to be M$_{\rm bary}$ $\simeq$ 8\,$\times$\,10$^{8}$ \msun.
At the last measured point of the \HI\ rotation curve (see
Figure~\ref{figcap9}), the implied interior dynamical mass is
$\sim$9\,$\times$\,10$^{9}$ \msun.  While we are sensitive to neither
the molecular component of the ISM nor to the cool dust component with
the present observations, recent investigations have shown that these
components are not substantial mass reservoirs in the interstellar
media of low-mass galaxies (see, e.g., {Leroy
  \etal\ 2005}\nocite{leroy05}, {Walter \etal\ 2007}\nocite{walter07},
{Schruba \etal\ 2012}\nocite{schruba12}, and various references
therein).  We conclude that UGCA\,105 is a typical dwarf galaxy whose
mass is dominated by dark matter (especially at large galactocentric
radii).  These results are similar to those found in recent studies of
other nearby dwarf galaxies (see, e.g., the detailed discussion in {Oh
  \etal\ 2008}\nocite{oh08}, {Cannon \etal\ 2010}\nocite{cannon10},
and {Oh \etal\ 2011}\nocite{oh11}).

\section{Conclusions}
\label{S4}

We have presented new low resolution {\it JVLA} imaging of the nearby
low mass galaxy UGCA\,105.  This system has remained comparatively
under-studied in the astrophysical literature.  The system is actively
forming stars (as traced by high surface brightness \halpha\ emission;
see {Kennicutt \etal\ 2008}\nocite{kennicutt08}) and is located in
sufficient proximity (3.39$\pm$0.25 Mpc; {Jacobs
  \etal\ 2009}\nocite{jacobs09}, {Tully \etal\ 2009}\nocite{tully09})
to allow detailed dynamical analysis.  In this work we present the
first spatially resolved study of the neutral gas dynamics of
UGCA\,105.

At 54\arcsec\ (890 pc) resolution, the neutral gas morphology and
kinematics are measured with high fidelity with these new {\it JVLA}
data. The \HI\ gas spans $\sim$175 \kms; sampled over $\sim$54
channels each separated by 3.3 \kms, the system displays a classical
``butterfly diagram'' and double-horned integrated line profile.  We
recover 98\% of the single dish flux \citep{springob05}, with no
correction for \HI\ self absorption applied.  We derive a systemic
velocity of 90.8\,$\pm$2.0 \kms; UGCA\,105 has an unusual velocity
given its location well outside the Local Group.

The integrated \HI\ column density distribution of UGCA\,105 contains
high surface density gas (N$_{\rm HI}$ $>$ 10$^{21}$ cm$^{-2}$)
throughout the extent of the luminous stellar disk.  While we see
evidence for regions of low column density (i.e., \HI\ holes or
shells) and also evidence for spatial agreement between regions of
ongoing star formation (as traced by \halpha\ emission) and high
column density gas, we defer a detailed treatment until higher spatial
resolution \HI\ imaging is available.  We do note that the inclination
corrected surface mass density profile of UGCA\,105 falls in the
regions closest to the dynamical center; this could be interpreted as
marginal evidence for a molecular region in the inner disk.

These new {\it JVLA} data offer an opportunity to study the bulk
neutral gas dynamics of UGCA\,105.  The system displays well-ordered
rotation throughout the neutral gas disk and the intensity weighted
isovelocity contours are parallel throughout the entire inner disk
(cospatial with the high surface brightness stellar disk).  We use
standard GIPSY tilted ring analysis in order to fit the observed
velocity field of UGCA\,105.  Regardless of how the parameters are
fixed, we find a robust and well constrained rotation curve at all
galactocentric radii.  The profile rises steeply in the innermost
$\sim$1.5 kpc, rises more slowly in the region from $\sim$1.5 kpc to
$\sim$5 kpc, and then remains flat at 72 \kms\ out to the last
measured point (7.5 kpc).  The total dynamical mass of UGCA\,105,
derived from this rotation curve, is M$_{\rm dyn}$ $=$
(9$\pm$2)\,$\times$\,10$^{9}$ \msun.  This dynamical mass is larger
than the sum of the luminous components [M$_{\rm gas}$ $=$
  (5.9$\pm$0.7)\,$\times$\,10$^{8}$ inclusive of a 35\% correction for
  Helium and molecular material; M$_{\star}$ $=$
  (1.8\,$\pm$\,0.8)\,$\times$\,10$^{8}$ \msun] by a factor of
$\sim$10; UGCA\,105 is a typical, dark matter dominated dwarf galaxy
(see, e.g., {Oh \etal\ 2008}\nocite{oh08}, {Oh
  \etal\ 2011}\nocite{oh11}, and references therein).

The proximity and favorable inclination (55\degr) of UGCA\,105 make it
a promising target for high resolution studies of both star formation
and rotational dynamics in a nearby low-mass galaxy.  In particular, B
configuration {\it JVLA} observations would achieve a synthesized
physical resolution element of order 100 pc; the high \HI\ surface
brightness would guarantee high signal to noise measurements at this
spatial resolution.  The resulting datasets would facilitate detailed
analyses of the interplay of neutral gas and recent star formation on
$\sim$100 pc scales.

\acknowledgements
 
The authors would like to thank the National Radio Astronomy
Observatory for making the ``Observing for University Classes''
program available to the astronomical community, and for organizing a
very enjoyable and productive visit to the Science Operations Center
in Socorro, New Mexico.  We acknowledge a helpful anonymous referee
whose suggestions helped to improve the quality of this manuscript.
J.M.C. would like to acknowledge useful discussions with Bradley
Jacobs, Igor Karachentsev, Robert C. Kennicutt, Jr., and Janice
Lee. J.M.C. would like to thank Paul Overvoorde, Daniel Hornbach,
Kathleen Murray, Ann Minnick, and Tonnis ter Veldhuis, whose
contributions to this project allowed it to be a success.  Finally,
the authors would like to thank Macalester College for generous
research and teaching support.

\bibliographystyle{apj}                                                 


\end{document}